\documentclass[twocolumn]{aastex62}

\usepackage{times}
\usepackage{array}
\newcolumntype{P}[1]{>{\centering\arraybackslash}p{#1}}
\usepackage[T1]{fontenc}

\graphicspath{{./}{figures/}}

\shorttitle{ALMA Solar System Analogs}
\shortauthors{Long et al.}

\usepackage{graphicx}
\usepackage{subfigure}
\usepackage{amsmath}
\usepackage{multirow} 
\usepackage{gensymb}

\begin{document}

\title{Hints of a Population of Solar System Analog Planets from ALMA}

\author[0000-0003-3840-7490]{Deryl E. Long}
\affil{Department of Astronomy, University of Michigan, 1085 S. University Avenue, Ann Arbor, MI 48109, USA}

\author[0000-0002-0661-7517]{Ke Zhang}
\affil{Department of Astronomy, University of Michigan, 1085 S. University Avenue, Ann Arbor, MI 48109, USA}
\affiliation{Hubble Fellow}

\author[0000-0003-1534-5186]{Richard Teague}
\affil{Center for Astrophysics | Harvard \& Smithsonian, 60 Garden Street, Cambridge, MA 02138, USA}

\author[0000-0003-4179-6394]{Edwin A. Bergin}
\affil{Department of Astronomy, University of Michigan, 1085 S. University Avenue, Ann Arbor, MI 48109, USA}

\begin{abstract}

The recent ALMA DSHARP survey provided illuminating results on the diversity of substructures in planet forming disks. These substructures trace pebble-sized grains accumulated at local pressure maxima, possibly due to planet-disk interactions or other planet formation processes. DSHARP sources are heavily biased to large and massive disks that only represent the high (dust flux) tail end of the disk population. Thus it is unclear whether similar substructures and corresponding physical processes also occur in the majority of disks which are fainter and more compact. Here we explore the presence and characteristics of features in a compact disk around GQ Lup A, the effective radius of which is 1.5 to 10 times smaller than those of DSHARP disks. We present our analysis of ALMA 1.3mm continuum observations of the GQ Lup system. By fitting visibility profiles of the continuum emission, we find substructures including a gap at $\sim$ 10 au. The compact disk around GQ Lup exhibits similar substructures to those in the DSHARP sample, suggesting that mechanisms of trapping pebble-sized grains are at work in small disks as well. Characteristics of the feature at $\sim$ 10 au, if due to a hidden planet, are evidence of planet formation at Saturnian distances. Our results hint at a rich world of substructures to be identified within the common population of compact disks, and subsequently a population of solar system analogs within these disks. Such study is critical to understanding the formation mechanisms and planet populations in the majority of protoplanetary disks. 

\end{abstract}

\keywords{ planetary systems: protoplanetary disks -- circumstellar matter --  planets and satellites -- planet-disk interactions}

\section{Introduction} \label{sec:intro}

Much exciting work has been focused on determining the link between protoplanetary disk properties and the resulting populations of planets \citep{Benz_2014}. As the diverse population of known exoplanets continues to grow, we are compelled to better understand the origins of these worlds as well as those of our own solar system. Planetesimal formation, a critical step in the formation of terrestrial worlds and giant planet cores, remains somewhat of a challenge, as  radial drift could remove disk solids ultimately by accretion of sublimated grain material by the star \citep{whipple_1972, WEIDEN_1997}. However, if there are pressure maxima at specific locations in the disk they might stop radial drift and perhaps foster growth as pebbles accumulate \citep{Lyra_2008, Johansen_2009, Pinilla_2012, Teague_2018}.  These regions could be isolated by searching for substructure, such as gaps and/or rings in submm/mm continuum emission from disk solids \citep[e.g.,][]{Dullemond_2018}.  Thus, searching for these substructures is a crucial component of understanding the formation processes and presence of planets in circumstellar disks \citep{Andrews_2020}. 

Beyond grain growth, there is now substantial work on observable features in disks as being induced by forming planets \citep{Wolf_2005, Dodson_Robinson_2011, Zhu_2011, Gonzalez, Pinilla_2012, Ataiee, Bae_2016, Kanagawa_2016, Rosotti_2016, Isella_2018}, and characteristics of these substructures, such as gap width and depth, have been used to infer properties of the potential forming planets \citep{Kanagawa_2015,Kanagawa_2016, Dong_2017}. 

The Disk Substructures at High Angular Resolution Project (DSHARP) conducted using ALMA provided high resolution data of 20 large, bright disks. The results of DSHARP reveal a diverse and abundant population of substructures, which have been linked to planet formation through subsequent study \citep{Zhang_2018}. While this initial survey, and others such as the survey of disks in the Taurus star-forming region by \citet{Long_2018}, provides compelling observational results linking circumstellar disk properties to planet formation, it is biased toward large disks with high dust flux- only a small portion of the full disk population. We are left wondering if, and to what extent, substructures and corresponding formation processes also occur in smaller disks. 

\begin{figure}
\begin{center}
\includegraphics[width=\columnwidth]{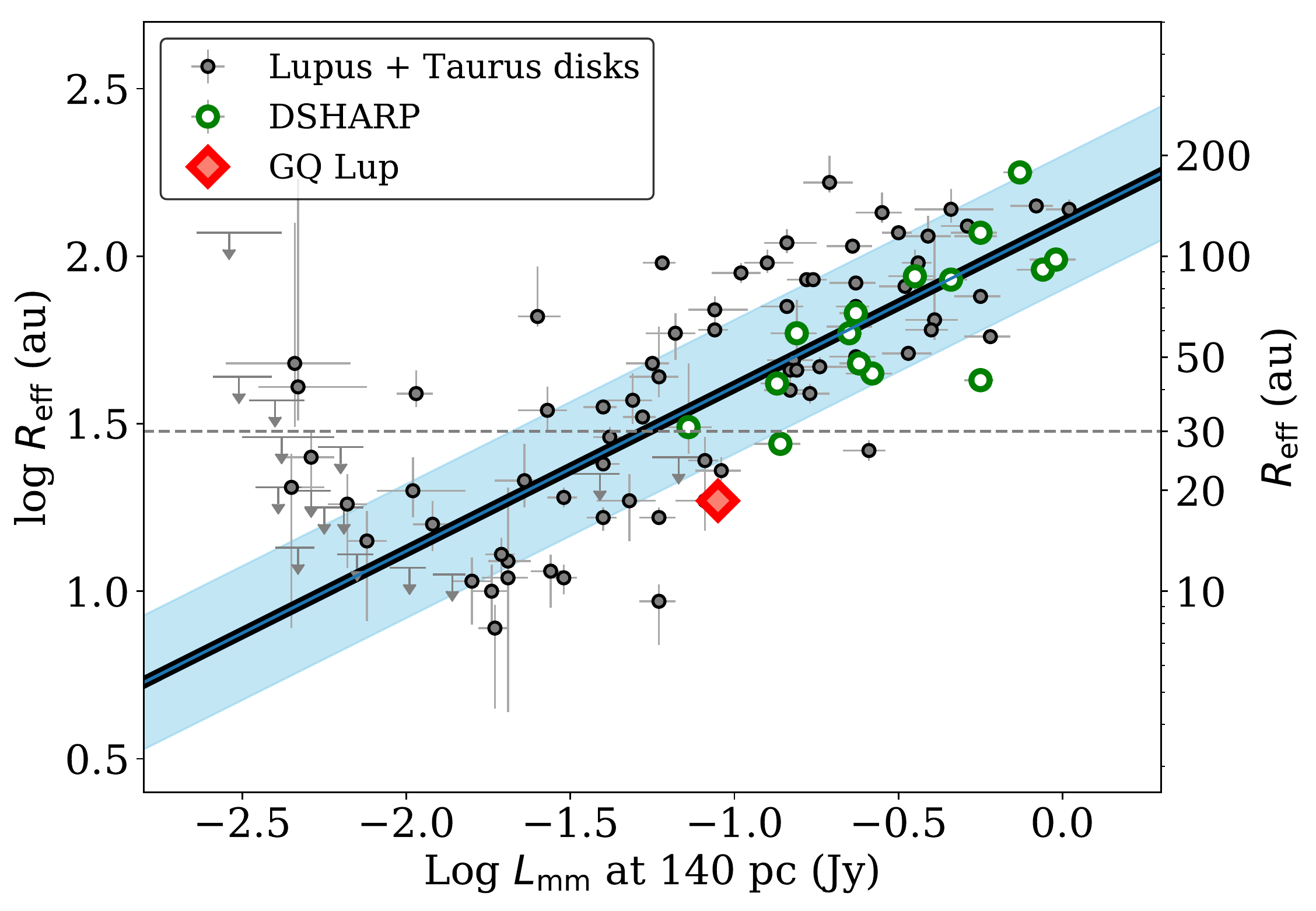}
\caption{Size-luminosity relationship for millimeter continuum sources with disk properties from \citet{Andrews_2018}. We show GQ Lup (red diamond) in comparison to Lupus and Taurus sources (grey circles) as well as DSHARP sources (green circles). It shows that GQ Lup moves toward a discovery space of smaller disks. \label{fig:1}}
\end{center}
\end{figure}

The majority of disks are faint and exhibit compact (sub)mm continuum emission with effective radii $<$ 20 au \citep{Ansdell_2016, Pascucci, Williams_2019}. Effective radius (R$_\text{eff}$) is defined as the radius at which the cumulative flux encompasses 68 percent of the total disk continuum flux \citep{Andrews_2018}. While R$_\text{eff}$ is not a measure of a disk's outer cut-off, and one does see emission and substructure outside of R$_\text{eff}$, it is used here to demonstrate GQ Lup's disk size relative to those in DSHARP and other studies. Compact disks present a useful population of study in the pursuit of understanding the formation processes of solar system analog planets. High angular resolution observations with ALMA make this study possible, allowing us to resolve substructures in the dust continuum. We utilize 1.3 mm observations from ALMA to study GQ Lup, a source with a compact disk. In Figure \ref{fig:1}, we show that GQ Lup is much more compact than the DSHARP sources, with an effective radius of 19 au as compared to the average DSHARP size of 50 au \citep{Andrews_2018_DS}. It is an important step forward to show that ALMA can be used to accomplish investigations of planet formation in small disks. Using ALMA data of GQ Lup at 1.3mm, we explore the presence of substructures and the potential of hidden planets, including one at Saturnian distances. This work moves toward an ALMA discovery space of solar system analog planets forming in compact disks. 

\section{Methodology} \label{sec:methods}

\subsection{Observations and Data Reduction}
This work utilizes ALMA observations of the GQ Lup System. Continuum analysis was performed using the $\#$2015.1.00773.S. data set first presented by \citet{Wu_2017}. These observations were taken on UT
2015, November 1 in Cycle 3 with the Band 6 receiver. The observations utilized 41 12-m antennas for a  total on-source time of 11 minutes. The baseline coverage was 84.7\textendash14969.3 meters. The three basebands configured for continuum observations are centered at 233.0, 246.0, and 248.0 GHz respectively, and provide a total continuum bandwidth of 2 GHz. Data were reduced using the CASA package (version 5.0.0-218) \citep{McMullin}. We followed data reduction steps outlined in \citet{Wu_2017}, performing three rounds of phase-only self calibration with the TCLEAN algorithm with natural weighting, giving us a synthesized beam of 0\farcs{057} $\times$ 0\farcs{032}. We ensure that the image is properly centered by fitting a Gaussian to the continuum source and centering the image using \textit{fixvis} in CASA. Our continuum map was produced by imaging the calibrated measurement set with the TCLEAN algorithm, and has an rms of 56 ${\mu}$Jy beam$^{-1}$. From imaging our calibrated data set with all continuum windows, we retrieve an integrated flux of 26 mJy for the GQ Lup A disk, well matching results from \citet{Wu_2017}.

\section{Results \& Analysis} \label{sec:analysis}
\subsection{Deriving the surface brightness profile of the 1.3 mm continuum}
To uncover substructures in the GQ Lup A disk, we employ an empirical model-fitting approach in the visibility domain \citep{Pearson}. This approach has the advantage of using full baseline lengths and thus can recover smaller spatial scale structures than the CLEAN images \citep{Zhang_2016}.  

The data are deprojected using an inclination angle $i = 60\degree$ and a position angle $\phi = 346\degree$ \citep{MacGregor_2017}. A source distance of 152 pc \citep{gaia_2018} is used for conversions between angular distance and au. The deprojected visibilities $(u',v')$ and radial brightness profile $(I_{v}(\theta))$ are related by a Hankel transform \citep{Pearson}: 

\begin{equation}
    u' = (u\cos{\phi} - v\sin{\phi}) \times \cos{i}
\end{equation}

\begin{equation}
    v' = u\sin{\phi} + v\cos{\phi}
\end{equation}

\begin{equation}
    V(\rho) = 2\pi \int_{0}^{\infty} I_{v}(\theta) \theta J_{0}(2\pi\rho\theta) d\theta
\end{equation}

\noindent where $\rho = \sqrt{u'^{2} + v'^{2}}$ is the deprojected \textit{uv}-distance in units of $\lambda$, $\theta$ is the radial angular scale from the center of the disk, and $J_{0}$ is a Bessel function.

We model $I(\theta)$, the disk surface intensity distribution corresponding to intensity in the image domain, with a parametric function (Equation \ref{eq4}) developed by \citet{Zhang_2016}. The key feature of the model function is that it is a composite of Gaussian functions modulated with sinusoidal amplitudes.

\begin{equation} \label{eq4}
\begin{split}
    I(\theta) = & \frac{a_{0}}{\sqrt{2\pi}\sigma_{0}}\exp \left(-\frac{\theta^{2}}{2\sigma_{0}^{2}}\right) \\ & + \sum_{i} \cos(2\pi\theta\rho_{i}) \times \frac{a_{i}}{\sqrt{2\pi}\sigma_{i}}\exp \left(-\frac{\theta^{2}}{2\sigma_{i}^{2}}\right)
\end{split}
\end{equation}

This parametric model has been tested with ALMA observations of several protoplanetary disks and shown robust matches to long-baseline observations \citep[e.g.][]{ADAM_2018}. 

\begin{table*}[]
\centering
    \begin{tabular}{ |P{1.5cm}||P{2.7cm}|P{2.7cm}|P{2.7cm}|}
     \hline
     \multicolumn{4}{|c|}{Best-Fit Parameters} \\
     \hline
    Gaussian & $a$ (Jy/arcsec) & $\sigma$ (arcsec) & $\rho$ (k$\lambda$) \\
    
     \hline
     1   &  $0.129\pm{0.0012}$  &  $0.0777\pm{0.0006}$ & 0.0 (fixed)  \\
     2 &  $0.0338\pm{0.0029}$  &  $0.0791\pm{0.0061}$ & $1615\pm{28}$  \\

     \hline
     \end{tabular}
    \caption{Optimal values of free parameters and uncertainties from our fitting using MPFIT and the parametric model of disk surface intensity distribution from \citet{Zhang_2016}. These values describe the two Gaussians which are used to model the deprojected visibility profile (Figure \ref{fig:2}). 
    \label{table:1}}
\end{table*}

Model-fitting was performed using MPFIT \citep{mpfit}, a Levenberg-Marquardt least-squares minimization algorithm. This routine iteratively fits our parametric model to the real visibilities by searching for optimum values of free parameters $\{a_{0},\sigma_{0}, a_{i}, \sigma_{i}, \rho_{i}\}$. We provide initial guesses for the amplitudes $\{a_{i}\}$ and central locations of Gaussians $\{\rho_{i}\}$ based on peaks in the visibility profile. We use two Gaussians based on the two peaks seen in the visibility profile, though we note that this model underestimates the uv profile at the shortest baseline. A model with three Gaussians provides the best fit at the shortest baseline, as shown in the left panel of Figure \ref{fig:2}, and suggests additional substructures may exist in GQ Lup A. However, given that there are only two clear peaks in the visibility profile, we proceed with a best-fit parametric model using two Gaussians: one for the central peak and one for the peak around 1600\,k$\lambda$. We report best-fit parameters and errors in Table \ref{table:1}. The real visibilities, best-fit model, and model radial intensity profile are shown in Figure \ref{fig:2}.

We produce 2D images based on the best-fitting radial profile and compare directly to the observed continuum image (Figure \ref{fig:3}). In general, the best-fit model does an excellent job in reproducing the observed continuum image. Since our best-fit model is axisymmetric, the most significant residuals are asymmetric features. We find 7$\sigma$ residuals, but the current data do not have sufficient resolutions to identify the detailed shape of these asymmetric features. For following analysis, we focus only on axisymmetric substructures.

\subsection{Characterizing Substructures}
  
Using our best-fit model of the radial surface brightness profile, we identify substructure features (e.g., gaps and rings) by searching for local maxima and minima. We employ the same method that \citet{Huang_2018} used to identify substructures in the DSHARP sample, labeling dark gap features with the prefix D and bright ring features with the prefix B. Figure 12 in \citet{Huang_2018} shows a schematic for their definition of width, which we employ here.

\begin{figure*}
\begin{center}
\includegraphics[width=\textwidth]{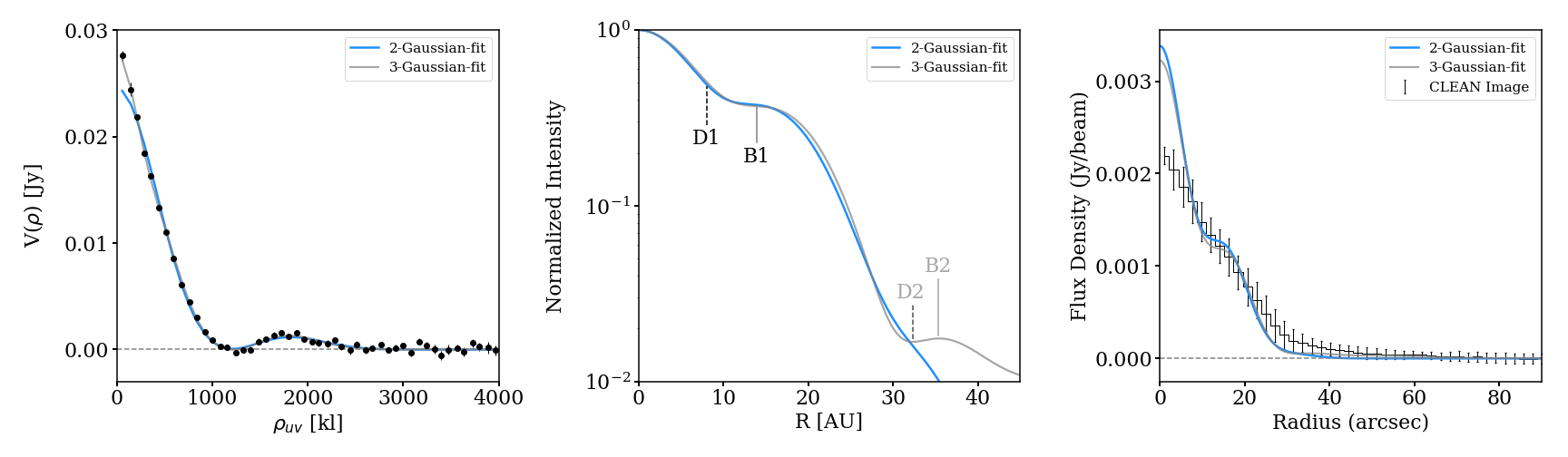}
\caption{ (left) Visibility profile for continuum emission of GQ Lup. Black dots show the real parts of visibilities. In each panel, the blue curve is our best-fitting model, while the grey curve shows a 3-Gaussian-fit that best fits the data at the shortest baselines but lacks motivation from peaks in the visibility profile. (center) Our model radial intensity profile for GQ Lup A. Dashed lines denote gaps and solid lines denote rings. (right) Azimuthally averaged radial profile of the observed flux density (CLEAN image) compared with the model prediction from visibility-fitting in blue. The visibility-fitting method generally matches the CLEANed method but also reveals smaller-scale structures. 
\label{fig:2}}
\end{center}
\end{figure*}

\begin{figure*}
\begin{center}
\includegraphics[width=\textwidth]{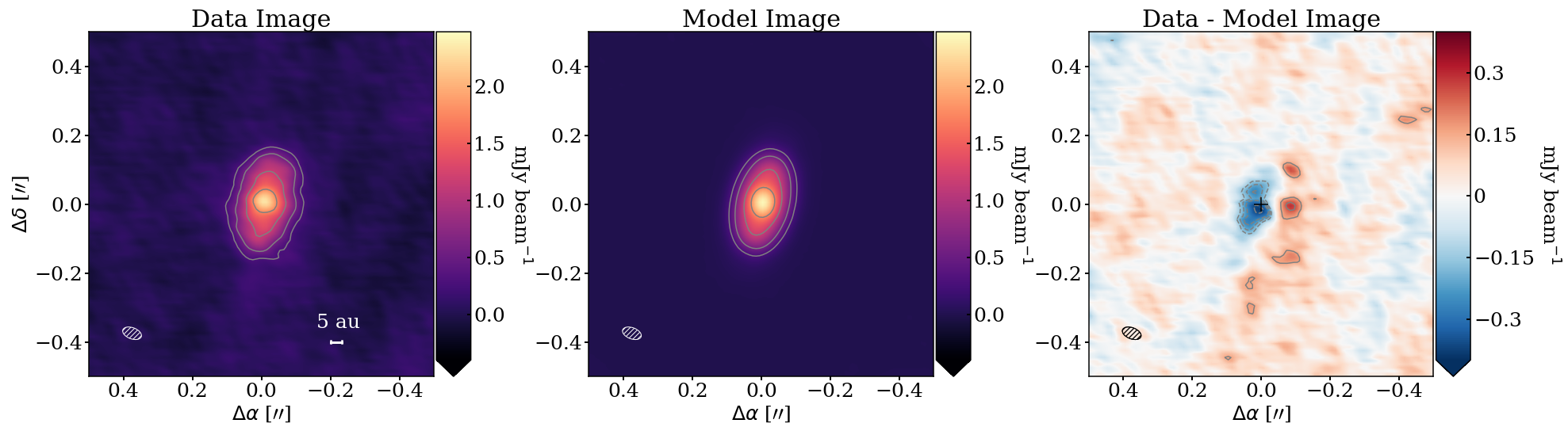}
\caption{ Our data image, model image, and residuals. For the data and model image, contour levels are in steps of [5, 10, 20, 30] x $\sigma$ (where $\sigma$ = 56 ${\mu}$Jy beam$^{-1}$, the rms of the image). For our residuals we plot contour levels [-7, -5, -4, -3, 3, 5] x $\sigma$ ($\sigma$ = 52 ${\mu}$Jy beam$^{-1}$). Negative values are dashed contours and positive values are solid. \label{fig:3}}
\end{center}
\end{figure*}

Some substructures do not exhibit a clear minimum or maximum, but rather a ``plateau" shape, while still demonstrating a deviation from the regular disk profile. We see this for the feature centered at $\sim$ 10 au (Figure \ref{fig:2}). \citet{Huang_2018} find similar features in MY Lup and DoAr 33. We numerically compute the gradient of the radial intensity profile in the region of interest and determine where the change in radial intensity becomes significantly small, using the condition that $ \frac{1}{I_{v}(r)} \frac{dI_{v}(r)}{dr}$ be greater than -0.05 \citep{Huang_2018}. 

For D1/B1 we lack a robust method for measuring depth, as the intensity $I_{b}$ of the brighter region where the gap ends is inherently lower than the intensity $I_{d}$ where the gap begins. To estimate the depth, we fit a function to the data on either side of the feature, extending 5 au on either side where $ \frac{1}{I_{v}(r)} \frac{dI_{v}(r)}{dr}$ is less than -0.05 and directly compare the intensity in the region of the plateau feature to the intensity in the same region given by the function.

Using these methods we identify and characterize a robust feature at $\sim$ 10 au, as well as a tentative feature at 32 au which is suggested by the 3-Gaussian-fit model. The gap at 32 au (D2\footnote[1]{The gap D2 is presented tentatively because it is only suggested by the 3-Gaussian-fit model and not by the best-fit 2-Gaussian model. Values associated with this tentative feature are reported but are denoted with an asterisk (e.g., D2*).}) has a width of 2.25 au and a depth of 0.957. Its neighboring ring at 35 au (B2*) has a width of 3.15 au. The gap at 8 au (D1) is more broad in shape and its center falls within the range 8-13 au. The center panel of Figure \ref{fig:2} shows the locations of identified substructures, and values associated with the two gaps are listed in Table \ref{table:2}.

\subsection{Masses of Hidden Planets}

Here we interpret gaps in the GQ Lup disk in the context of disk-planet interactions. However, we note that the origins of substructures in disks are still under debate. Other possible origins include changes in dust properties near condensation fronts of dominant ices \citep{zhang_2015, oku_2016}, and magneto-hydrodynamic processes \citep{flock_2015}. Low-contrast structures in the radial profile could also be related to temperature variations rather than gas density variations alone. Breaking the degeneracy between temperature and density requires observations of multiple CO transitions \citep{facchini_17}. Kinematic information from line observations can further confirm the presence of planets \citep{Teague_2018, teague_2019}. 

To estimate the masses of inferred planets we begin with a total dust mass of 5.9$M_{\Earth}$ from Model A in \citet{Wu_2017}, one of their best-fit models. We convert this value to total gas mass using a gas-to-dust ratio of 100. We also use a second, lower value of total gas mass that \citet{MacGregor_2017} retrieved using a simple parametric model of disk structure. Given the large difference between values in gas mass for these methods, we use both in order to provide a range of planet mass estimates.

We approximate the total gas mass as the total disk mass and convert from total disk mass to $\Sigma_{0}$, gas surface density at 1 au, using

\begin{equation}
    m_{d} = \frac{2\pi\Sigma_{0}}{2 - \gamma}(r_{out}^{2-\gamma}-r_{in}^{2-\gamma}),
\end{equation}

\noindent with radius values $r_{in}$ = 1.5 au and $r_{out}$ = 23.8 au from Model A in \citet{Wu_2017}. We solve for $\Sigma_{g,0}$, the gas surface density at the location ($r_{planet}$) of a possible planet, using the relation \begin{equation}
  \Sigma{(R)} = \Sigma_{0}\Big(\frac{R}{1 \: au}\Big)^{-\gamma}  .
\end{equation}

We employ the model used by \citet{Wu_2017}, with $\gamma$ = 0.1, which has a fairly flat disk surface density distribution. We also test steeper distributions using $\gamma$ = 1 and $\gamma$ = 1.5, but proceed with $\gamma$ = 0.1 as all of the distributions lead to the same choices moving forward. 

An initial disk gas mass of 590 $M_{\Earth}$ gives us a gas surface density of approximately 10 g cm$^{-2}$ at the locations of gaps D1 and D2*, while an initial disk gas mass of 71.3 $M_{\Earth}$ gives us a gas surface density of approximately 1 g cm$^{-2}$ at these locations. Using

\begin{equation}
  St = 1.57\times10^{-3} \frac{\rho_{p}}{1 g cm^{-3}} \frac{s}{1 mm} \frac{100 g cm^{-2}}{\Sigma_{g,0}} , 
\end{equation}

\noindent and assuming a maximum particle size of 0.1 mm (p = 3.5 and s$_{max}$ = 0.1) \citep{Hull_2018, Kataoka_2017}, we calculate Stokes numbers of 1.57$\times$10$^{-3}$ and 1.57$\times$10$^{-2}$, respectively. Using these values we select related models from Tables 1 and 2 in \citet{Zhang_2018} to perform our subsequent calculations of planet-star mass ratios. We use fitting relationships outlined by \citet{Zhang_2018} to relate gap width and depth to the planet-star mass ratio, $q$. \citet{Zhang_2018} use power laws to fit observable quantities, defining a parameter K as being proportional to $q$ and having a power law dependence on scale height. Least squares fitting is used to determine coefficients (A, B, C, and D) for relationships between depth and its optimal degeneracy parameter K, as well as between width and its optimal degeneracy parameter K'.

For width-fitting we use $\Delta$ values determined from the inner and outer radii of the gap features listed in Table \ref{table:2}. $\Delta = (r_{out} - r_{in})/r_{out} $.  We solve for K' and determine the planet mass ratio $q$ using 

\begin{equation}
   \Delta = AK'^{B}, 
\end{equation}

\noindent and

\begin{equation}
    K' = q(h/r)^{-0.18}\alpha^{-0.31}. 
\end{equation}
 
We proceed with depth-fitting using $\delta = I_{b} / I_{d} $, the inverse of the depth values listed in Table \ref{table:2}. We solve for K and determine the planet mass ratio $q$ using
 
 \begin{equation}
  \delta - 1 = CK^{D},  
\end{equation}

\noindent and

 \begin{equation}
K = q(h/r)^{-2.81}\alpha^{-0.38} .    
\end{equation}

  To produce their suite of models, \citet{Zhang_2018} used three different scale height values, $(h/r)$ = (0.05, 0.07, 0.1), and three different values for the disk turbulent viscosity coefficient, $\alpha$ = ($10^{-4}, 10^{-3}, 10^{-2}$). We choose the middle values, simulating ``average" conditions. We convert from planet-star mass ratios, $q = M_{p}$/$M_{*}$, to planet mass estimates using the mass of GQ Lup A, 1.05$M_{\odot}$ \citep{Wu_2017} and present our estimates in Table \ref{table:2}.

The upper mass values for a potential planet in feature D1 are large as this feature has a ``plateau" shape. Lacking a robust method for measuring its width, we employ the same definitions as \citet{Zhang_2018}. We further consider that broad, shallow gaps may be a sign of low viscosity (small $\alpha$), as is the case for AS 209 \citep{Fedele_2018}. We repeat the process using $\alpha$ = $10^{-4}$ assuming a lower viscosity disk and include the resulting planet mass values in Table \ref{table:2} for comparison. We note that there is a wide range of mass values for this feature due to difficulties characterizing plateau features. 

We ultimately present a range of mass estimates for inferred planets in the gap at 8 au, and tentative gap at 32 au. For a low viscosity disk ($\alpha$ = $10^{-4}$) we present a mass range of 3.07-355 $M_{\Earth}$ for D1 and a range of 0.62-2.44 $M_{\Earth}$ for D2*. For a more viscous disk ($\alpha$ = $10^{-3}$) we present a mass range of 7.37-724 $M_{\Earth}$ for D1 and a range of 1.27-5.86 $M_{\Earth}$ for D2*.

\begin{table*}[t]
\centering
    \begin{tabular}{ |p{1.0cm}|p{1.0cm}|p{1.5cm}|p{1.5cm}|| p{2cm}|p{1.5cm}|p{1.7cm} | p{1.5cm}|p{1.5cm}|}
     \hline
     \multicolumn{9}{|c|}{Gaps and Inferred Planet Masses} \\
     \hline
    \multirow{2}{*}{Feature}& \multirow{2}{*}{$r_{0}$ (au)} & \multirow{2}{*}{Width (au)}& \multirow{2}{*}{Depth}  & \multirow{2}{*}{Fitting Method} & \multirow{2}{*}{ $m_{d}$ ($M_{\Earth}$)} & \multirow{2}{*}{q ($M_{p}$/$M_{*}$)} & \multicolumn{2}{c|}{Planet Mass $M_{\Earth}$ }\\
    \cline{8-9}
      &  &  & & & & & $\alpha$ = $10^{-3}$ & $\alpha$ = $10^{-4}$  \\
           \hline
        \multirow{4}{*}{D1} & \multirow{4}{*}{8} & \multirow{4}{*}{5.85} & \multirow{4}{*}{0.940} & \multirow{2}{*}{Width} & 590 & 2.08 x $10^{-3}$ & 724 & 355 \\ 
                            &                    &                   &                      &                   &71.3 &  9.59 x $10^{-4}$ & 334 & 164\\ \cline{5-9}
                            &                    &                   &                      &  \multirow{2}{*}{Depth} & 590 &  2.12 x $10^{-5}$  &  7.37 & 3.22\\ 
                            &                    &                   &                      &                   & 71.3  & 2.22 x $10^{-5}$  & 7.73 &  3.07  \\ \hline
                            
        \multirow{4}{*}{D2*} & \multirow{4}{*}{32} & \multirow{4}{*}{2.25} & \multirow{4}{*}{0.957} & \multirow{2}{*}{Width} & 590 & 5.00 x $10^{-5}$ & 1.27 & 0.62\\ 
                            &                    &                   &                      &                   & 71.3 &  5.76 x $10^{-6}$ & 2.01 & 0.98\\ \cline{5-9}
                            &                    &                   &                      & \multirow{2}{*}{Depth} & 590 &   1.66 x $10^{-5} $ &  5.77 & 2.40 \\ 
                            &                    &                   &                      &                   & 71.3  & 1.68 x $10^{-5}$  & 5.86  & 2.44  \\
     \hline
     \end{tabular}
    \caption{Radial locations and characteristics of gaps and corresponding mass estimates for inferred planets. We report mass estimates for features D1 and D2*, which show the most promise of containing a hidden planet. We present values for both width and depth fitting techniques \citep{Zhang_2018}, which are broken down further  based on initial disk mass ($m_{d}$) and $\alpha$. Our scaled mass value of 590 $M_{\Earth}$ is based on gas-to-dust ratio of 100 and a dust mass of 5.9 $M_{\Earth}$ \citep{Wu_2017}, and 71.3 $M_{\Earth}$ is the gas mass value given by \citet{MacGregor_2017}. 
    \label{table:2}}
\end{table*}

\section{Discussion and Conclusions} \label{sec:discussion}

This work presents the detection of dust substructures in mm emission of a faint, compact disk. Using high resolution ALMA 1.3mm observations and modeling in the visibility domain, we detect substructures in the GQ Lup disk. Our study of GQ Lup contributes to an effort to build a comprehensive understanding of disk evolution and planetary formation, which balances the rich high-resolution observations of bright disks with equally rich studies of faint disks. 

\subsection{Origins of Compact Disks}

The study of substructures in compact disks has implications for our understandings of disk formation and evolution. In their ALMA study of compact disks in the Taurus star-forming region, \citet{Long_2019} did not see significant substructures in continuum images for compact disks ($<$ 50 au) at 0.1" resolution. Such compact disks may have lost their outer dust disk over time due to rapid radial drift, with no trapping mechanisms at work to stop it. This effect is seen in the case of CX Tau, where the large difference in extent between continuum emission and molecular emission indicates strong radial drift \citep{Facchini}. 

Alternatively, compact disks might simply be born small or may be truncated due to tidal effects of companions in multiple systems. \citet{Manara} find that the majority of disks around primary stars in multiple systems have dust radii that are significantly smaller than disks around single objects. Since GQ Lup A has a companion at a projected separation of $\sim$ 110 au, its disk evolution may have been affected by interactions with its companion. High resolution observations of CO emission in GQ Lup and additional compact sources will be necessary to understand the affects of truncation and build a larger picture of the physical mechanisms occurring in compact sources.

Our detection of substructures in a compact disk suggest that pressure gradients and dust traps are present within smaller disks. We argue that low-contrast or narrow unresolved substructures are present in many compact disks, perhaps including those in the Taurus sample studied by \citet{Long_2019}. These pressure gradients likely prevent rapid inward radial drift from depleting the reservoir of pre-planetary solids. Further study of substructures in compact disks in both multiple and single systems will provide important insights and constraints to the evolution and planet-forming abilities of the larger population of compact disks.  

\subsection{Origins of Substructures}

Understanding how disk environments contribute to planet formation offers insight to the diversity of planet properties we observe, as well as the formation of our own solar system. There is a growing interest in inferring the properties of young forming planets from the dust substructures in a given disk. The substructures we detect in the GQ Lup disk may be induced by young forming planets. If induced by a forming planet, the feature centered at $\sim$ 10 au is evidence of planet formation at Saturnian distances. The compact disk therefore represents an environment in which we may study the formation of solar system analogs. 
 
 We provide mass ranges for planets which may be forming in the GQ Lup disk. Planets inferred to be forming in the larger DSHARP disks \citep{Zhang_2018} are generally Neptune mass planets past 10 au. The planets we infer in the GQ Lup disk are possibly smaller, particularly in the tentative gap D2* at radius 32 au which has a planet mass range 0.62-5.86 $M_{\Earth}$. As for the broader feature D1, we present a wider and higher planet mass range than that of D2*. We find that planet mass is higher at smaller radii, though cannot make generalizations as we are limited to one source. Our findings are consistent with understandings of giant planet formation which show that it is rare to find giant planets on wide orbits \citep{Zhang_2018, Bowler_2018}. 
 
 We note again that GQ Lup has a substellar companion. It has been suggested that substructures in binary systems may be caused by tidal interactions with a companion. \citet{Wagner_2018} show that the spiral arm features in the disk around HD 100453A may be related to companion interaction rather than embedded planets or other causes. We must therefore consider that substructures in the GQ Lup disk may be related to tidal interactions with GQ Lup b. We do not yet see clear spiral structure in the disk, and rich information about the system configuration and binary orbit would be necessary to either confirm or rule out tidal interactions as a cause of substructures in the GQ Lup disk. Thus we currently present embedded planets as a possible cause for the gap features identified in this work.  
 
 Our preliminary study of GQ Lup hints that a rich world of substructures are present in small disks, which may be related to young forming planets. We present the detection of substructures and hints of planet formation at Saturnian distances in the disk of GQ Lup and conclude that compact disks may represent a space in which to probe the formation of solar system analogs. Further high resolution study of disk substructures in compact disks is necessary in order to infer the broader population of young forming planets in these environments. 

\vspace{1em}

We thank the anonymous referee for their insightful comments, which improved the quality of this paper. This work makes use of the following ALMA data:  ADS/JAO.ALMA $\#$2015.1.00773.S. ALMA is a partnership of ESO (representing its member states), NSF (USA) and NINS (Japan), together with NRC (Canada), NSC and ASIAA (Taiwan), and KASI (Republic of Korea), in cooperation with the Republic of Chile. The Joint ALMA Observatory is operated by ESO, AUI/NRAO and NAOJ. The National Radio Astronomy Observatory is a facility of the National Science Foundation operated under cooperative agreement by Associated Universities, Inc. Results from distributed computing were obtained using the Chameleon testbed
supported by the National Science Foundation. We acknowledge that ALMA operates on indigenous Atacame\~no (Likan-antai) land. K.Z. acknowledges the support of NASA through Hubble Fellowship grant HST-HF2-51401.001 awarded by the Space Telescope Science Institute, which is operated by the Association of Universities for Research in Astronomy, Inc., for NASA, under contract NAS5-26555.

\bibliographystyle{aasjournal}
\bibliography{gqlup}

\end{document}